\documentclass{article}    

\usepackage{graphicx}

\title{\textbf{Without the Born Rule}}  
\author{Richard Mould\footnote{Department of Physics and Astronomy, State University of New York, Stony Brook,
\mbox{New York} 11794-3800; http://ms.cc.sunysb.edu/\~{}rmould}}  
\date{}    
   
\begin{document}             

\maketitle              

\begin{abstract}

The auxiliary rules of quantum mechanics have always included the ``Born rule" that connects probability with square modulus. 
This need not be the case, for it is possible to introduce probability into the theory through probability current alone.  When
this is done, other rules can provide for stochastically triggered measurements within a system of any size,
microscopic or macroscopic; and solutions to the Schr\"{o}dinger equation can be consistently applied to individual trials, not
just to ensembles of trials.  Other advantages appear.  The rules can then resolve the paradox associated with the Schr\"{o}dinger
cat experiment, and remove the possibility of the many world thesis of Everett.  As a result, the system can accommodate any
conscious observer, including the principal investigator who cannot otherwise be included in a quantum mechanical system.  

\end{abstract}

\section*{Introduction}
	Schr\"{o}dinger's equation is always accompanied by a set of auxiliary rules that say how it is to be used.  The Copenhagen school
gives one set of rules, four of which are listed in another paper \cite{RM1}.  They include the Born rule, instructions relating
to the use of primary data, and the collapse of the wave function, etc.  Other rule-sets of standard quantum mechanics have been
proposed in the past 80 years, some of which deny that there is a collapse of the wave function.   However, they all have one thing
in common.  They all include the Born rule that functions as the sole connection between theory and observation.  I will call any
one of these rule-sets the \emph{sRules}, indicating that they are auxiliary rules that accompany standard quantum mechanics.  

	The Born rule is not necessary.  It is possible to achieve very desirable results by dropping the Born rule from the governing
auxiliary rules and introducing probability into quantum mechanics through \emph{probability current} alone.  This has been done
with two very different auxiliary rule-sets called the \emph{nRules} and the \emph{oRules}.  In this paper we are only concerned
with the nRules.

\section*{The nRules}
	For the most part, solutions to Schr\"{o}dinger's equation represent a continuous and classical-like evolution of a quantum
mechanical wave function.  Occasionally this evolution engages a state that is discontinuous with its immediate predecessor in some
variable.  This may be called a quantum \emph{jump} or quantum \emph{gap}.  The nRules are concerned with what happens at one of
these gaps.  

We choose the nRules with two objectives in mind.  First, the rules should locate any measurement  inside of the system being
measured, and should apply equally to all systems independent of size -- microscopic or macroscopic.  A measurement is understood to
mean a stochastically initiated collapse of the wave that occurs independent of an `external' measuring device or observer. 
Second, the rules should apply to individual trials, not just to ensembles of trials.  When these objectives are met, we find that
the system is able to include conscious observers without the ambiguity associated with the Schr\"{o}dinger cat experiment, or
with the many worlds of Everett.

	The first of these rules recognizes two different kinds of states: \emph{ready states} and \emph{realized states}.  A ready state 
$\underline{S}$ is underlined, and a realized state $S$ is not.  The first rule tells us that ready states are introduced into
solutions of Schr\"{o}dinger's equation immediately after a quantum jump.  A state is otherwise realized.  We define a \emph{ready
component} as one that includes a ready state.  It too exists immediately after a quantum jump.

\vspace{.4cm}

\noindent
\textbf{nRule (1)}: \emph{If an irreversible interaction produces a complete component that is discontinuous with its predecessor
in some variable, then it is a ready component.}

\vspace{.4cm}

Complete components contain all of the symmetrized objects in the universe.  Each included object is itself complete in that it is
not just a partial expansion in some representation.  The ready states in a discontinuous component are the discontinuous states
appearing there.  

The second rule establishes the existence of a stochastic trigger that stochastically chooses ready components.   The flow per
unit time of square modulus between components is given by the modular current $J$, and the total square modulus of  the system is
given by $s$.  The square modulus is not here identified with probability; rather, probability is introduced only through
\emph{probability current} $J/s$.  The division of $J$ by $s$ automatically normalizes the system at each moment of time.  So
currents rather than functions are normalized in this treatment.

\vspace{.4cm}

\noindent
\textbf{nRule (2)}: \emph{A systemic stochastic trigger strikes a ready component with a probability per unit time equal to the
positive probability current J/s flowing into it.}

\vspace{.4cm}

The collapse of a wave function and the change of a ready state to a realized state is given by nRule (3).

\vspace{.4cm}

\noindent
\textbf{nRule (3)}: \emph{When a ready component is stochastically chosen, all the included ready states become realized states,
and all other components go to zero.}

\vspace{.4cm}

A complete statement of the first three rules is given in Appendix A together with other relevant definitions and needed
clarifications.  There is a fourth nRule.  However, we will illustrate the first three rules before going on.

\section*{Particle and Detector Interaction}
	We apply Schr\"{o}dinger's equation to a `microscopic' particle interacting with a 'macroscopic' detector in order to demonstrate
the first three nRules.  These two objects are assumed to be initially independent and given by the equation 
\begin{equation}
\Phi(t) = exp(-iHt)\psi_i\otimes d_i
\end{equation}
where $\psi_i$ is the initial particle state and $d_i$ is the initial detector state.  The particle is then allowed to pass over the
detector, where the two interact with a cross section that may or may not result in a capture.  After the interaction begins at a
time $t_0$, the state is an entanglement in which the particle variables and the detector variables are inseparable.  In general,
a direct product of states such as $\psi d$ (in Eq.\ 2 below) indicates an interaction that, in most cases, leads to or has led
to an entanglement. 
                                                 
The first component in Eq.\ 2 includes the detector $d_0$ in its ground state prior to capture.  The second component
$\underline{d}_1$ is the detector in its capture mode.
\begin{equation}
\Phi(t \ge t_0) = \psi_id_0(t) + \underline{d}_1(t)
\end{equation}
where the capture state is zero at $t_0$ and increases in time.  It is a `ready' state as prescribed by nRule (1), for the
particle goes discontinuously and irreversibility from being spread throughout space in the first component to being inside the
detector in the second component -- like going discontinuously from the $1^{st}$ to the $2^{nd}$ orbit of an atom.

	The state $\psi(t)$ is an incoming free particle plus all of the scattered components that are correlated with recoil states of
the ground state detector.   The second component is a superposition that includes all of the recoil components of the
detector that have captured the particle\footnote{Each component in Eq.\ 2 has an attached environmental $E_0$ and $E_1$ that is
not shown.  These are orthogonal to one another, insuring local decoherence.  But even though Eq.\ 2 may be decoherent locally, we
assume that the macroscopic states $d_0$ and $d_1$ are fully coherent when $E_0$ and $E_1$ are included.  So Eq.\ 2 and others
like it in this paper are understood to be coherent when universally considered. We call them ``superpositions", reflecting their
global rather than their local properties.  Superpositions of mesoscopic states have been found at low temperature \cite{JRF}. 
The components of these states are locally coherent for a measurable period of time.}.

	As probability current flows into the second component in Eq.\ 2, it may take a stochastic hit at some time $t_{sc}$ following
nRule (2).  In that case the state will collapse following nRule (3) to yield
\begin{displaymath}
\Phi(t \ge t_{sc} > t_0) = d_1(t)
\end{displaymath}
where $d_1$ becomes a realized state and the first component disappears.  Renormalization is not necessary because the current flow
into $\underline{d}_1(t)$ is continuously normalized according to nRule (2).  It is explained in Ref.\ 1 how to treat this case when
there is no capture of the particle.  

So far there are no difficulties with these nRules; however, difficulties arise when there are
three or more components as in the case of a particle counter.

\section*{A Counter}
	If a particle counter is exposed to a radioactive source its state function would normally be written
\begin{equation}
\Phi(t \ge t_0) = C_0(t) + C_1(t) + C_2(t) + C_3(t) + etc.
\end{equation}
where $C_0$ is the initial state of the counter (registering 0 counts) at time $t_0$.  The state $C_1$ registers 1 count, $C_2$
registers 2 counts, etc.  The radioactive source and the particle field are not shown in Eq.\ 3.  The
components following $C_0$ are zero at $t_0$; and after that, they become non-zero in the form of a pulse that travels from left
to right in Eq.\ 3.    

	Here is the difficulty.  We want the nRules to apply to macroscopic systems as well as microscopic systems.  If the interaction
is a long-lived radioactive decay, then the pulse in Eq.\ 3 might be spread over two or more components  at the
same time.  It might overlap both $C_1$ and $C_2$, causing probability current to flow simultaneously into each one.  This means
that there might be a stochastic hit on $C_2$ \emph{before} there is a hit on $C_1$, and that would cause a collapse of the state
that goes directly to $C_2$.  The state $C_1$ would then be skipped over.  This is a very unphysical result for a macroscopic
body, and the fourth nRule is designed to insure that that does not happen.

\section*{The Fourth nRule}
Only positive current going into a ready component is physically meaningful because it represents positive probability.  A negative
current (coming out of a ready component) is not physically meaningful and is not allowed by nRule (4).

\vspace{.4cm}

\noindent
\textbf{nRule (4)}: \emph{A ready component cannot transmit probability current to other components or advance its
own evolution.}

\vspace{.4cm}

Although it can receive current that increases its square modulus, a ready state is dynamically terminal.  It cannot develop beyond
itself or contribute to the development of anything else.  It is shown in the Appendix A that this property is arranged by
modifying the Hamiltonian of the system.  When that is done the first term $C_0(t)$ in Eq.\ 3 decreases in time as the second
component $C_1(t)$ increases in time, conserving the square modulus; but $C_1(t)$ no longer evolves dynamically beyond the values
given it by $C_0(t)$ at each moment of time.  The correct nRule equation is therefore
\begin{equation}
\Phi(t \ge t_0) = C_0(t) + \underline{C}_1(t)
\end{equation}
replacing Eq.\ 3.  This insures that $\underline{C}_1$ \emph{will be chosen}.  It will not be skipped over.

	There is no general theory that determines the Hamiltonian associated with Eq.\ 4.  In standard theory it is found from long
experience to take the form $H_s = H_0 + H_{01} + H_1$, where $H_0$ and $H_1$ drive the first and second components respectively,
and $H_{01}$ drives the interaction between them.  In the end, the justification for this form is that it works.  However, nRule (4)
forces a  change.  The Hamiltonian now takes the truncated form $H_0 + H_{01}$, where the last term $H_1$ has been
dropped (see Appendix A).  This too works, but it works differently.  It works when combined with the other nRules.  The effect of
this change is demonstrated in the remainder of this paper.  

Think of $\underline{C}_1(t)$ in Eq.\ 4 as the \emph{launch component} of a next solution of Schr\"{o}dinger's equation if and
when it is stochastically chosen.  It contains all the boundary conditions of that solution.  Those boundary conditions are
carried over from $C_0$ directly into $\underline{C}_1$ at each moment of time (through $H_{01}$), but they are not applied to the
next solution of the Schr\"{o}dinger equation until the moment of collapse to that solution.  If there is a stochastic hit on
$\underline{C}_1$ at time
$t_{sc}$, then nRule (3) will require a collapse to the next solution that is the realized \mbox{state $C_1$}.
\begin{equation}
\Phi(t \ge t_{sc} > t_0) = C_1(t) + \underline{C}_2(t)\hspace{.5cm}\mbox{now using} \hspace{.5cm}H_1 + H_{12}
\end{equation}
where the ready state $\underline{C}_2(t)$ is the launch component into the `next' solution of the Schr\"{o}dinger equation.  

	We see that the \emph{single} equation given by the sRules of standard quantum mechanics in Eq.\ 3 is replaced by a separate
equation for each quantum jump, plus one.  So the first quantum jump in Eq.\ 3 is replaced by \emph{two} equations under the
nRules -- one \emph{before} the stochastic hit (Eq.\ 4), and one \emph{after} the stochastic hit (Eq.\ 5).  This is characteristic
of the difference between the sRules and the nRules.  Whereas the sRules allow only one solution beginning with the initial
conditions, the nRules take on new boundary conditions and hence new solutions to the Schr\"{o}dinger equation after each
stochastic hit.  The sRules run all these solutions together as though they all occur at the same time.  The nRules take them one
at a time in their proper sequence.  

	Since the nRules are intended to govern microscopic as well as macroscopic systems, equations like 4  and 5 should apply as well
to atomic states.  If the atom goes through a series of decays from levels $A_5(t)$ to $A_2(t)$ and then to the ground level
$A_0(t)$, it will do so in  separate steps. 
\begin{eqnarray}
\Phi(t \ge t_0) = A_5(t) + \underline{A}_2(t)&\mbox{using}& H_5 + H_{52} \\
\mbox{followed by}\hspace{.3cm}\Phi(t \ge t_{sc1} >t_0) = A_2(t) + \underline{A}_0(t)&\mbox{using}& H_2 + H_{20}\nonumber\\
\mbox{and then}\hspace{.9cm}\Phi(t \ge t_{sc2} > t_{sc1} > t_0) = A_0(t)&\mbox{using}& H_0\nonumber
\end{eqnarray}
where $t_{sc1}$ and $t_{sc2}$ are the times of the two stochastic hits that initiate the two decays. Starting with the initial
conditions $A_5(t_0)$, the first solution in \mbox{Eq. 6} prepares the launch state $\underline{A}_2(t)$ of the second solution
which contains all of the required boundary conditions of $A_2(t)$ at the time  of the first decay.  The second solution prepares
the launch state $\underline{A}_0(t)$ of the final solution which contains all of the required boundary conditions of $A_0(t)$ at
the time of the second decay.  In each case the launch component is not ``launched" until it is stochastically chosen, for a
ready component cannot evolve dynamically on its own.    

On the other hand, the sRules will run these solutions together into one equation.  Experimentally, the ensemble of photons of
either frequency that is received during any interval of time after $t_0$ is the same for the nRules as is for the sRules.

\section*{Parallel Branching}
Now imagine parallel states in which a quantum process may go either clockwise or counterclockwise as shown in Fig.\ 1.  Each
component includes a macroscopic piece of laboratory apparatus $A$, where the Hamiltonian provides for an irreversible and
discontinuous clockwise interaction going from the $0^{th}$ to the $r^{th}$ state and another one from there to the final state
$f$; as well as a comparable counterclockwise interaction from the $0^{th}$ to the $l^{th}$ state and from there to the final
state $f$.  The Hamiltonian does not provide a direct route from the $0^{th}$ to the final state, so the system chooses
stochastically between a clockwise and a counterclockwise route.  Launch states $\underline{A}_l$ and $\underline{A}_r$ are the
eigenstates of that choice, and contain the boundary conditions of each choice.

\begin{figure}[h]
\centering
\includegraphics[scale=0.8]{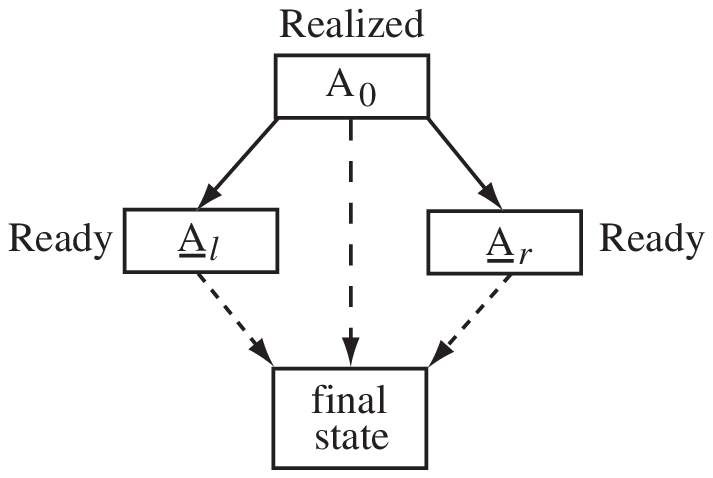}
\center{Figure 1: Possible parallel decay routes}
\end{figure}
With nRule (4) in place, probability current cannot initially flow from either one of the intermediate states to the final state,
for that would require current flow out of a ready state.  The dashed lines in Fig.\ 1 indicate the forbidden transitions.  But
once the state $\underline{A}_l$ (or $\underline{A}_r$) has been stochastically chosen, it will become a realized state $A_l$ (or
$A_r$) and the final state will become $\underline{A}_f$.  A subsequent transition to $\underline{A}_f$ can occur that      
realizes $A_f$.

The effect of nRule (4) is therefore to force this macroscopic system into a classical sequence that goes either clockwise or
counterclockwise.  Without it, the system might make a second order transition (through one of the intermediate states) to the
final state without the intermediate state being realized.  The observer might then see the initial state followed by the final
state, without knowing (in principle) which pathway was followed.  This is a  property of quantum mechanical systems that applies
to continuous microscopic processes under the nRules as well as the sRules.   In Heisenberg's famous example formalized by
Feynman, a microscopic particle observed at point $a$ and later at point $b$ will travel over a quantum mechanical superposition
of all possible paths in between.  Without nRule (4), macroscopic objects facing discontinuous and irreversible parallel choices
would do the same thing.  But that should not occur.  The fourth nRule forces this parallel system into one or the other path, so
it is not a quantum mechanical superposition of both paths.

The same would be true of a microscopic parallel system of discontinuous paths.  Suppose $A_0$ is an initial atomic energy level
where there are two separate paths to the ground state $A_f$.  The sRules say that these two paths are in superposition.  Everett's
many-world theory furthermore tells us that the two paths will not be further involved with one another -- except in this case,
since they arrive at the same final state.  The nRules tell a different story.  They deny that these two paths are in
superposition.  They have no intermediate involvement with one another other \emph{because} only one of them happens at a time. 
ItÕs a classical either/or situation, even at this atomic level.

\section*{Free Neutron Decay}
Another interesting example is that of a free neutron decay.  This is written
\begin{displaymath}
\Phi(t) = n(t) + \underline{ep\overline{\nu}}(t)
\end{displaymath}
where the second component is zero at $t = 0$ and increases in time as probability current flows into it.  This component contains
three entangled particles that make a whole object, where the three together are a single ready component as indicated by the
underline (see nRule 1).  Following nRules (2) and (3), there will be a stochastic hit on $\underline{ep\overline{\nu}}(t)$ at some
time
$t_{sc}$, reducing the system to the realized correlated states $ ep\overline{\nu}(t_{sc})$. 

This case provides a good example of how $\underline{ep\overline{\nu}}(t)$ is a function of time beyond its increase in square
modulus.  Assume that the neutron moves across the laboratory in a wave packet of finite width.  At each moment the ready
component $\underline{ep\overline{\nu}}(t)$ will ride along with the packet, with the same size, shape, and group
velocity.  It is the launch component that contains the boundary conditions of the next solution of the Schr\"{o}dinger equation --
the one that appears when $\underline{ep\overline{\nu}}(t)$ is stochastically chosen at time $t_{sc}$.  Before this `collapse',
$\underline{ep\overline{\nu}}(t)$  is time dependent because it increases in square modulus \emph{and} because it follows the
motion of the neutron.  However, nRule (4) insures that it will not evolve dynamically beyond itself until it becomes a realized
component at the time of stochastic choice.  The neutron $n(t)$ will then disappear and the separate particles in the launched
component $ep\overline{\nu}(t)$ will spread out on their own, although correlated in conserved quantities.

Specific values of, say, the electron's momentum are not stochastically chosen by this reduction because all possible values of
momentum for each particle are included in $ep\overline{\nu}(t)$ and its subsequent evolution (after $\Delta E$  is resolved). 
For the electron's momentum to be determined in a specific direction away from the decay site, a detector in that direction must be
activated.  That will require another stochastic hit on, and collapse to, the component that includes that detector.

\section*{Spin Correlated Fermions}
Still another example is a pair of correlated fermions $f_1f_2$, where the direct product generally assumes an interaction that
leads to, or has led to, an entanglement of some kind.  It is not necessary to be specific about this entanglement in order to
write the minimal equation (under the nRules) of a measurement of the spin of the second particle.  This is given by
\begin{displaymath}
\Phi(t \ge t_0) = f_1f_2M + \underline{F}_1(\uparrow)\underline{M}(\downarrow) +
\underline{F}_1(\downarrow)\underline{M}(\uparrow)
\end{displaymath}
where the measuring device $\underline{M}(\downarrow)$ now includes the second fermion and records a spin down, and the measuring
device $\underline{M}(\uparrow)$ includes the second fermion and records a spin up.  In general, the $FM$ products might also be
entanglements.  Current flows from the first component to both of these launch components.  If there is a stochastic hit on the
second component at time $t_{sc}$ we will have
\begin{displaymath}
\Phi(t \ge t_{sc} > t_0) = F_1(\uparrow)M(\downarrow)
\end{displaymath}

\section*{Add an Observer}
	Assume that an observer is looking at the detector in Eq.\ 1 from the beginning.  
\begin{equation}
\Phi(t) = exp(-iHt)]\psi_i\otimes D_iB_i
\end{equation}
where $B_i$ is the observer's initial brain state that interacts with the detector $D_i$.  The brain is understood to include only
higher order brain parts -- that is, the physiology of the brain that is directly associated with consciousness after all image
processing is complete.  All lower order physiology leading to $B_i$ is assumed to be part of the detector.  The detector is now
represented by a capital $D$ indicating that it includes the bare detector plus the low-level physiology of the observer.  

	Following the interaction between the particle and the detector, Eq.\ 2 becomes
\begin{equation}
\Phi(t \ge t_0) = \psi_i(t)D_0(t)B_0 + \underline{D}_{1w}(t)B_0
\end{equation}
where the launch component is zero at $t_0$ and increases in time. The state $B_0$ in the first component is the observer's
conscious brain state that interacts with the detector $D_0(t)$, and is therefore aware of its `zero' reading.  The launch state
$D_{1w}(t)$ represents the capture state of the detector when only its window end had been affected.  When the particle enters the
window of the detector it initiates an irreversible process that defines a ready component, and then it freezes because of nRule
(4). This means that except for the particle's presence just inside the window, the launch component in Eq.\ 8 has the \emph{same
content} as the first component.  Therefore, $B_0$ is the brain state that appears in the launch component.  Physically, this
reflects the fact that the observer at the display end will not experience the capture until its effect has moved classically
through the detector to the display, and that does not happen until \emph{after} a stochastic hit \mbox{on
$\underline{D}_{1w}(t)$}. 

With that hit on $\underline{D}_{1w}(t)$ at time $t_{sc}$, the system becomes

\begin{equation}
\Phi(t \ge t_{sc} >t_0) = D_{1w}(t)B_0 \rightarrow D_{1d}(t)B_1
\end{equation}
where the realized component $D_{1w}(t)B_0$ evolves continuously and classically (represented by the arrow) until it reaches the
display end of the detector (represented by $D_{1d}$).  At this point the brain state $B_1$ of the observer becomes aware of the
capture.  The change in the conscious brain state from $B_0$ to $B_1$ in Eq.\ 9 is continuous and classical.

	Compare this with an sRule evaluation of this capture.  In that case Eq.\ 8 would be given by
\begin{displaymath}
\Phi(t \ge t_0) = \psi(t)D_0(t)B_0 + D_{1w}(t)B_0 \rightarrow D_{1d}(t)B_1\hspace{1cm}\mbox{(standard QM)}
\end{displaymath}
where only $\psi(t)D_0(t)B_0$ is initially non-zero.  The plus sign again refers to a discontinuous quantum jump and the arrow
refers to a continuous classical evolution.  So as probability current builds up the second (window) component, current will begin
to flow from the window to the display state in a continuous evolution.  If the particle in this equation is produced by a
long-lived radioactive decay, this single equation might very well contain both $B_0$ and $B_1$ in superposition.  The observer
would then be in two different states of consciousness \emph{at the same time}.  Here we have another formulation of
Schr\"{o}dinger's famous cat paradox, which is an unacceptable ambiguity.  

The nRules tell a different story.  According to the nRules there are two different equations in Eqs.\ 8 and 9.  One describes
the system \emph{before} the stochastic hit, and the other describes the system \emph{after} the stochastic hit.  Furthermore,
there is only one brain state in Eq.\ 8, and there is only one brain state `at a time' in \mbox{Eq.\ 9}, so there is \emph{no}
paradoxical cat-like ambiguity.  Again, the sRules run both solutions together as though they both occur at the same time, giving
rise to the ambiguity; whereas the nRules provide separate solutions that apply before and after the stochastic hit, eliminating
the ambiguity.  Each of these separate solutions has its own boundary conditions.  The initial conditions apply to \mbox{Eq.\ 8},
and the boundary conditions created by the launch state in \mbox{Eq.\ 8} apply to \mbox{Eq.\ 9}.

\section*{Many Parallel Sequences}
\begin{figure}[b]
\centering
\includegraphics[scale=0.8]{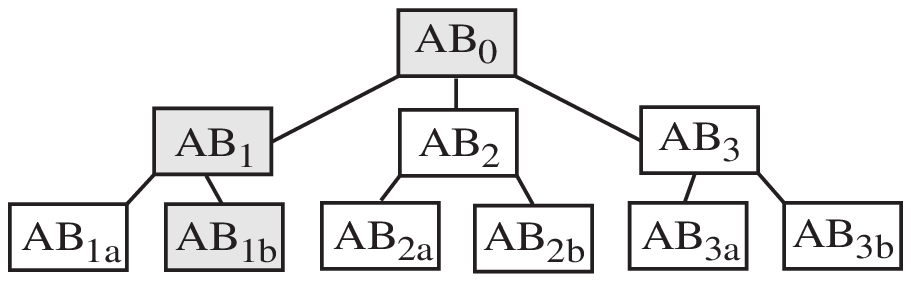}
\center{Figure 2: Six possible sequences}
\end{figure}
	Imagine three counters that surround a single radioactive source.  The system state $A_0$ means that none of the counters have yet
recorded a count.   The system state $A_1$ means the first counter is the first to record a count, $A_2$ means that the second
counter is the first to record a count, etc.  Each of the three counters is exposed to two other radioactive sources $a$ and $b$
that will become active after the first choice has been made.   The system state $A_{xa}$ (or $A_{xb}$) indicates which of
these two sources produces a second count of the $A_x$ counter.  An observer is on hand to determine which of the six possible
sequences is realized in an individual trial.  These sequences are shown in Fig.\ 2, where $AB_0$ means that the brain $B$
observers the system in the state $A_0$, etc.  In writing the component this way, we overlook the detail associated with the
classical evolution that occurs inside the apparatus.  We take only the end result of the observer interacting with the apparatus. 
The shaded sequence in Fig.\ 2 is the one discussed below.

The sRules run all these possibilities together in a single superposition.  This makes sense if the superposition refers only to an
ensemble.  But the theory would not then apply to individual trials, and this is a theoretical limitation we want to avoid.  On the
other hand, if the superposition is applied to individual trials, the result is again paradoxical.  It means that the single
observer would experience all possible sequences at once.  This is the \emph{many world} theses of Everett, according to which the
observer in Fig.\ 2 splits up into six separate conscious entities.  Everett shows that this does not lead to a contradictory
confrontation between any two of the alter-egos because any sequence, once begun, will proceed without any further involvement with
any other sequence \cite{HE}.  

The nRules resolve this ambiguity.  After $t_0$ and before there has been a hit on any counter, the system is given by
\begin{equation}
\Phi(t \ge t_0) = A_0(t)B_0 + \underline{A}_1(t)B_1 + \underline{A}_2(t)B_2 + \underline{A}_3(t)B_3 
\end{equation}
where initially only $A_0(t)B_0$ is non-zero.  This is a parallel system in which current flows to all three eigen-components
$\underline{A}_1(t)B_1$,  $\underline{A}_2(t)B_2$, and $\underline{A}_3(t)B_3$ that are the possible launch states of the next
solution\footnote{The brain state in the launch components of Eq.\ 10 should be ready states according to nRule (1).  But if the
internal classical evolution is considered, ready brains states would not exist.  The notation in Eq.\ 10 reflects the latter
consideration.}.  

After the first stochastic hit at time $t_{sc1}$, imagine that the chosen  reduced state is
\begin{displaymath}
\Phi(t \ge t_{sc1} > t_0) = A_1(t)B_1 + \underline{A}_{1a}(t)B_{1a} + \underline{A}_{1b}(t)B_{1b} 
\end{displaymath}
where only $A_1(t)B_1$ is initially non-zero.  The realized component in this equation is equal to the second component in the
shaded sequence in Fig.\ 2.  Again, we overlook the classical evolution that occurs inside the counter in this equation, so $B_x$ is
shown together with $A_x$.

A second hit gives the final state
\begin{displaymath}
\Phi(t \ge t_{sc2} > t_{sc1} > t_0) = A_{1a}(t)B_{1b}  
\end{displaymath}
where the realized component in this case is the third shaded component in \mbox{Fig. 2}, and where again, the classical evolution
inside the detector is overlooked.  

	The nRules do not support a superposition of all the sequential possibilities.  Instead, they produce only one sequence in a
single trial that reflects the stochastic choices that have been made in that trial. So the nRules apply to individual trials
\emph{and} they avoid the unphysical many-world interpretation of Everett.  The different sequences, once begun, have no further
involvement with one another \emph{because} they occur separately as classical either/or choices.

\section*{Other Cases}
	In Ref.\ 1, the nRules are applied to a number of other cases with good results.  It is shown that the Born rule follows as a
theorem when the nRules are applied to a terminal observation in a typical physics experiment.  The nRules also work well when the
observer does not initially interact with a detector (as in Eq.\ 7), but comes on board \emph{during} the primary interaction
between the particle and the detector.  The nRules also work as they should when a second observer is brought on board to share an
observation with the first observer.  

	Also in Ref.\ 1, several other microscopic systems are analyzed using the nRules.  We look at atomic emission and absorption,
separating out the influence of spontaneous vs. stimulated emission.    Also, Rabi oscillations and decoherence are
investigated.  It is shown that both, by themselves, avoid creating ready states and stochastic reduction.  We also examine the
effect of magnetic fields on spin states.  It is shown that there are no stochastic reductions prior to the spin system
interacting with a `location' detector after it has being deflected by the field.   

	In another paper \cite{RM2}, the nRules are applied the Schr\"{o}dinger cat experiment in its various forms.  In the first version
of that experiment, the cat is initially conscious and is rendered unconscious  as a result of a
 beta decay.  In the second version, the cat is initially unconscious and is awaken by an alarm that is stochastically
triggered by a beta decay.  In a final version, the cat initially unconscious and is awakened by an internal alarm that is in
competition with the above external alarm.  In all these cases, the nRules successful predict the expected experience
of the cat.  In addition, the nRules successfully predict the experience of an external observer who opens the box (containing the
cat) at any time during any one of the above experiments.

\section*{Observers Included}
	The Born rule is not only one of the sRules, but it is taken to be the sole connection between the theory and observation.  That
contact is instantaneous.  The Born rule observer can only `peek' at the system from time to time,
approaching a continuous experience only in a Zeno-like approximation.  The primary observer cannot himself be a continuous part
of the system under the sRules.  Furthermore, while the sRules allow a secondary observer (such as Schr\"{o}dinger's cat) into the
system in a continuous way, they predict that that observer can be in two different states of consciousness at the same time.  I
believe that to be a false prediction rather than an occasion to speculate about the existence of multiple alter-egos.  Physics
should construct theory to accommodate observable phenomena -- not invent `unobservable' phenomena to accommodate theory.

	On the other hand,  the nRules  make a place for the primary and the secondary observer.  The brain states
referred to in the above  were presumably those of a secondary observer.  There was never any ambiguity about what those
brains experienced at any moment of time in any one of the separate solutions; so quantum theory with these auxiliary rules
accurately predicts the experiences of any included observer.  This \emph{is} the connection between theory and observation under
the nRules.  It is the same as that theory/observation connection in classical physics.  Also as in classical
physics, the primary observer who is investigating a system can always look at a wider system that includes himself in a
continuous way.  A model of his own brain is then included in the system, correctly predicting his experience there.

\section*{The oRules}
	There is another rule-set that rejects the Born rule and relies on probability current to introduce probability into quantum
mechanics \cite{RM3, RM4}.  These are called the \emph{oRules}.  They differ from the nRules in that only brain states can be ready
states.  So only brains can be the basis states of a state reduction.  These rules reflect the notions of Wigner \cite{EW} and
von Neumann \cite{JvN} and others to the effect that the presence of a conscious observer is necessary to a state reduction. 
These rules seem to be every bit as adequate as the nRules, and I do not make a final choice between the two.  In the end we do
not really know what is required for the collapse of a wave function, so I believe that all of the consistent and observationally
accurate auxiliary-rule options should be on the table.

\section*{Appendix A}
\noindent
\textbf{nRule (1)}: \emph{If an irreversible interaction produces a complete component that is discontinuous with the initial
component of a solution, then all of the new states that appear in this component will be ready states.  All other states will be
realized.}

\hangindent=.2in
[\textbf{note:} A \emph{complete component} is a solution of Schr\"{o}dinger's equation that includes all of the (symmetrized)
objects in the universe.  It is made up of complete states of those objects, including all their state variables.  A component that
is a sum of less than the full range of a variable (such as a partial Fourier expansion) is not complete.]

\hangindent=.2in
[\textbf{note:} Continuous means continuous in all variables.  Although solutions to Schr\"{o}dinger's equation change continuously
in time, they can be \emph{discontinuous} in other variables -- e.g., the separation between the $n^{th}$ and the $(n + 1)^{th}$
orbit of an atom with no orbits in between.  A discontinuity can also exist between macroscopic states that are locally decoherent. 
For instance, the detector state $d_0$ (the ground state) and $d_1$ (the capture state) are discontinuous with respect to detector
variables.  There is no state in between.  Like atomic orbits, these detector states are considered to be `quantum jump'
apart.  The continuous `internal' evolution of a detector that follows a jump is not part of the jump.  The jump only occurs at
the \emph{window} end of a detector, and is complete as soon as something irreversible happens (inside the detector) to secure the
measurement.]

\hangindent=.2in
[\textbf{note:} The \emph{initial component} is the first complete component that appears in a given solution of Schr\"{o}dinger's
equation.  A solution is defined by a specific set of boundary conditions.  Boundary conditions change with the collapse of the
wave function.  The single component that survives a collapse will be complete, and will be the initial component of the new
solution.]

\vspace{.4cm}

\noindent
\textbf{nRule (2)}: \emph{For a system of total square modulus $s$ that has $n$ ready components, a stochastic trigger will choose
stochastically from among them.  The probability per unit time of such a choice among $m$ of theses components
at time $t$ is given by $(\Sigma_mJ_m)/s$, where the square modular current $J_m$ flowing into the $m^{th}$  component at
that time is positive.}

\hangindent=.2in
[\textbf{note:} Dividing $J$ by $s$ insures normalization at every moment of time.  There is no need to normalize after each state
reduction because currents, not functions, are normalized in this treatment.]

\vspace{.4cm}

\noindent
\textbf{nRule (3)}: \emph{If a ready component is stochastically chosen, then all of the included ready states will become
realized, and all other components in the superposition will be immediately reduced to zero.}

\hangindent=.2in
[\textbf{note:} The claim of an immediate (i.e., discontinuous) reduction is the simplest way to describe the collapse of the state
function.  A collapse is brought about by an instantaneous change in the boundary conditions of the Schr\"{o}dinger equation,
rather than by the introduction of a new `continuous' mechanism of some kind.] 

\hangindent=.2in
[\textbf{note:} This collapse does not preserve normalization.  That does not alter the probability of subsequent reductions
because of the way probability per unit time is defined in nRule (1); that is, current $J$ is divided by the total square modulus. 
Again, currents are normalized in this treatment, not functions.]

\section*{Appendix B}
Assume that a unitary operator $U(t)$ displaces the system in time.  An instantaneous translation through time is then given
by 
\begin{center}
$U(dt)= 1 - iHdt$\
\end{center}
or \hspace{2.3CM} $U(t + dt) = U(dt)U(t) = (1 - iHdt)U(t)$

\noindent
where $H$ is a Hermitian Hamiltonian.  Applying it to $\Phi(t)$ gives
\begin{equation}
\Phi(t + dt) = (1 - iHdt)\Phi(t) = \Phi(t) - iH\Phi(t)dt
\end{equation}
from which we get Schr\"{o}dinger's equation 
\begin{displaymath}
H\Phi(t) = (id/dt)\Phi(t)
\end{displaymath}
Conservation of probability current follows from this.  Also, operators that commute with this Hamiltonian are constants of the
motion.

	We then choose the Hamiltonian that drives the system  
\begin{displaymath}
\Phi(t) = S_1(t) + \underline{S}_2(t)
\end{displaymath}

According to Eq.\ 11 the change $\Phi(t + dt) - \Phi(t)$ in time $dt$ is equal to $Ð iH\Phi(t)dt$, so each term in the Hamiltonian
drives some part of the system.   The Hamiltonian of this system in standard quantum mechanics is $H = H_1 + H_{12} + H_2$, where
$H_1$ drives $S_1(t)$ by itself, $H_2$ drives $S_2(t)$ by itself, and $H_{12}$ drives the interaction between the two.  If we don't
want to drive $S_2(t)$ dynamically on its own, we simply remove that term in the Hamiltonian.  This gives
\begin{displaymath}
H = H_1 + H_{12}
\end{displaymath}
as the Hamiltonian that results when applying nRule (4).   Schr\"{o}dinger's equation is then 
\begin{equation}
H_1S_1(t) + H_{12}[S_1(t) + \underline{S}_2(t)] = (id/dt)\{S_1(t) + \underline{S}_2(t)\}
\end{equation}
The first term $H_1S_1(t)$ in this equation provides for the internal evolution of $S_1(t)$ by itself.  The second (square
bracketed) term provides for the changes in square modulus of the two components; and in addition, it transmits internal changes in
$S_1(t)$ to $\underline{S}_2(t)$.  So $\underline{S}_2(t)$ is a function of time because its square modulus changes in time, and
because the internal dynamical changes in $S_1(t)$ are continuously being transmitted to it.  It continuously updates the boundary
conditions of the new solution.  However, it does not evolve on its own because $H_2$ is not present.  These are the properties we
want for the launch state.  

Our major assumption is that there exists a unitary operator that displaces this system (with its truncated Hamiltonian) in time. 
That cannot be formally demonstrated.  It is justified only by Eq.\ 12 that appears to give the desired result.

\end{document}